\documentclass[journal=apchd5, manuscript=article]{achemso}
\DeclareUnicodeCharacter{2212}{-}
\DeclareUnicodeCharacter{2062}{}

\usepackage[T1]{fontenc}
\usepackage[utf8]{inputenc}
\usepackage{standalone}
\usepackage{textgreek}
\usepackage{transparent}
\usepackage{mathtools, nccmath}
\usepackage{tabularx,booktabs}
\usepackage{multicol}
\usepackage{tikz}
\usetikzlibrary{fit, matrix}
\usepackage{soul}
\usepackage{float}
\usepackage{lmodern}
\usepackage{xcolor}
\definecolor{MC}{RGB}{0, 105, 60} 
\usepackage[version=3]{mhchem} % Formula subscripts using \ce{}



\makeatletter
\let\l@addto@macro\relax
\makeatother
\usepackage[fontsize=9pt]{scrextend}
\AtBeginDocument{\singlespacing}
\author{Dmitriy~Yavorskiy}
\affiliation[1]{Institute of High Pressure Physics, Polish Academy of Sciences, Sokołowska 29/37, 01-142 Warsaw, Poland}
\alsoaffiliation[2]{Institute of Physics, Polish Academy of Sciences, Aleja Lotników 32/46, 02-668 Warszawa, Poland}
\alsoaffiliation[4]{CENTERA, CEZAMAT, Warsaw University of Technology, Poleczki 19, 02-822 Warsaw, Poland}

\author{Jan~Suffczyński}
\affiliation[5]{Institute of Experimental Physics, Faculty of Physics, University of Warsaw, Pasteura 5, 02-093 Warsaw, Poland}

\author{Rafał~Kowerdziej}
\affiliation[3]{Institute of Applied Physics, Military University of Technology, Kaliskiego 2, 00-908 Warsaw, Poland}

\author{Olga~Strzeżysz}
\affiliation[6]{Institute of Chemistry, Military University of Technology, Kaliskiego 2, 00-908 Warsaw, Poland}

\author{Jerzy~Wróbel}
\affiliation[2]{Institute of Physics, Polish Academy of Sciences, Aleja Lotników 32/46, 02-668 Warszawa, Poland}
\alsoaffiliation[3]{Institute of Applied Physics, Military University of Technology, Kaliskiego 2, 00-908 Warsaw, Poland}

\author{Wojciech~Knap}
\affiliation[1]{Institute of High Pressure Physics, Polish Academy of Sciences, Sokołowska 29/37, 01-142 Warsaw, Poland}
\alsoaffiliation[4]{CENTERA, CEZAMAT, Warsaw University of Technology, Poleczki 19, 02-822 Warsaw, Poland}

\author{Marcin~Białek}
\affiliation[1]{Institute of High Pressure Physics, Polish Academy of Sciences, Sokołowska 29/37, 01-142 Warsaw, Poland}

\email{marcin.bialek@unipress.waw.pl}
\title{Terahertz magnon-polaritons control using a tunable liquid crystal cavity}
\abbreviations{THz, AFMR, FP, NiO}
\begin{document}
%\begin{tocentry}
%\begin{figurehere}
%    \includegraphics[width=1.\textwidth]{tog.png}
%\end{figurehere}
%\end{tocentry}

\begin{abstract}
Strong coupling of light to a collective spin excitation in antiferromagnets gives rise to hybrid modes called magnon-polaritons.
They are highly promising for data manipulation and transfer at terahertz rates, much faster than in the case of ferromagnetic magnon-polaritons, which operate at GHz frequencies. Yet, control of terahertz magnon-polaritons by the voltage, i.e. without ohmic dissipation losses, remains challenging. Here, we showcase the ability to remotely control antiferromagnetic magnon-polaritons at room temperature using an electric field by integrating a highly birefringent liquid crystal layer into a terahertz Fabry–Perot cavity containing an antiferromagnetic crystal. Positioned several millimeters from the magnetic material, the liquid crystal allows for electrical manipulation of the cavity's photonic environment by control of its dielectric constant. This adjustment, in turn, influences the extent of magnon dressing by cavity photons, thereby controlling the vacuum Rabi oscillations of the magnon resonance coupled to a particular cavity mode. Our approach enables reversible tuning of magnon-photon hybridization that can be triggered without direct electrical contact or alteration of the magnetic medium. These findings pave the way for voltage-programmable terahertz magnonic devices and open new avenues for noninvasive control strategies in spin-based information processing technologies.
\end{abstract}

\twocolumn
\noindent
\section{Introduction}
Magnons are collective excitations of the electron spins in the magnetically ordered crystals. The ability to transmit information without a charge transfer, thus without Joule heating, makes them highly attractive to opto-spintronics. Antiferromagnetic, zone-centered ($k=0$) magnons have drawn immense attention recently, as they exhibit dynamics in the terahertz (THz) range, much faster than in the case of their ferromagnetic counterparts, characterized by the gigahertz (GHz) dynamics. A prerequisite for implementing antiferromagnetic magnons in opto-spintronics is their control by voltage bias, i.e., without electrical current flow. Such a type of manipulation of magnons at ambient conditions has been shown exclusively in the case of the multiferroic material BiFeO$_3$\cite{PRovillain_NM2010, AKumar_APL2011} so far. However, the development of a versatile method for voltage-bias control of THz magnons applicable for any type of insulating antiferromagnets at room temperature has been still a challenge.

In the strong light-matter coupling regime, vacuum Rabi oscillations, involving the periodic exchange of energy between matter and optical modes, overcome losses, resulting in the emergence of new hybrid modes. 
The strong coupling of light to quasiparticles, such as magnons, excitons, and plasmons, gives rise to polariton modes that can be controlled through their photonic components. In particular, the strong magnon-photon interaction leads to the formation of magnon-polaritons (MPs). These MPs exhibit both light- and matter-like properties, making them highly promising for next-generation high-speed information processing technologies. \cite{KRoux_NC2020, HYuan_PR2022} Over the past decade, cavity magnonics studies have been focused almost entirely on ferromagnetic magnons excited in the GHz range and tuned by the magnetic field.\cite{Abe_APL2011, Zhang_PRL2014, Tabuchi_PRL2014, Zhang_NC2015, KGrishunin_ACSPh2018, Everts_PRB2020, Potts_APL2020, Lachance-Quirion_S2020, Li_JAP2020, Bhoi_JAP2021, SGuo_NL2023} Only a few reports on modification of the GHz ferromagnetic MPs induced by the electric current\cite{JTHou_APL2024} or voltage bias\cite{Kaur:APL2016,Rao:APL2018} exist. The strong coupling to light of antiferromagnetic magnons in the THz range was demonstrated recently, mostly using temperature and magnetic field tuning.\cite{KGrishunin_ACSPh2018, PSivarajah_JAP2019, MBialek_PRB2020, MBialek_PRA2021, MBialek_APL2022, ABaydin:PRR2023, MBialek_PRA2023, TGHBlank_APL2023, TElijahKritzell_AOM2023, MBialek_AFM2024, JChen:APL2025} The magnetic field and temperature tuning of antiferromagnetic magnon-polaritons is either intrinsically inefficient or too slow from the view of practical applications. Despite being highly desired and crucial for applications in THz opto-spintronics, voltage bias manipulation of THz MPs has not yet been reported. 

Liquid crystals provide a highly versatile platform for THz photonic applications, including voltage-induced THz beam manipulation\cite{XFu_ACS2022, ZShen_APLP2024}, phase modulation\cite{OBuchnev_APL2013} and shifting,\cite{LWang_LSA2015} as well as the tunability of THz metamaterials.\cite{DShrekenhamer_PRL2013, RKowerdziej_APL2014, RKowerdziej_APL2015, NChikhi_SR2016, RKowerdziej_LC2016, GDENG_OME2021}

In this work, we demonstrate the remote electrical control of THz MPs using a liquid crystal cell. We achieve it by integrating, into a Fabry-Perot cavity, the liquid crystal cell and a $330\pm10$-$\mu$m-thick slab of nickel oxide (NiO), a room-temperature antiferromagnetic insulator. A separation between the antiferromagnetic layer and the liquid crystal is approximately 2.5~mm. By varying the voltage applied to the liquid crystal, we tune the spatial overlap of the optical modes of the Fabry-Perot cavity and the antiferromagnet layer (see Figs. Fig.~\ref{fig:LCC_scheme}(a) and Fig.~\ref{fig:LCC_scheme}(b)). This allows us to tune the interaction strength between the Fabry-Perot cavity mode and the magnon above room temperature without the need for external magnetic fields. The presented method of control of magnon-polaritons without direct electrical contact with the magnetic layer can be applied to a wide range of magnetic materials. As such, it is highly attractive for implementation in the THz opto-spintronic systems for control of information transfer.

\section{Results and Discussion}\label{sec2}
%\subsection{Experimental}

\begin{figure}[t!]
\centerline{\includegraphics[width=.95\linewidth]{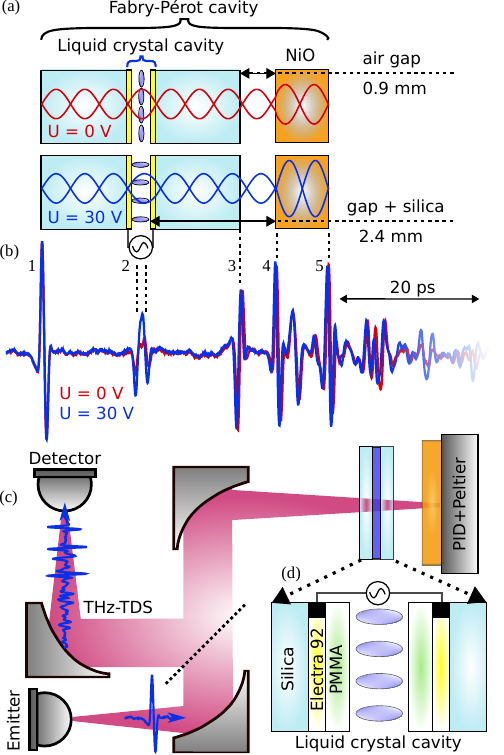}}
    \caption{(a) Schematic of the Fabry-Perot cavity incorporating a liquid crystal cell and an antiferromagnetic NiO crystal.
    (b) Time-domain reflection traces from the Fabry-Perot cavity for two selected values of the voltage bias.
    (c) Schematic of the quasi-optical setup.
    (d) Schematic of the liquid crystal cell, with electrodes made of conducting polymer Electra 92 highlighted in yellow and layers of the PMMA resist in green.
    }
    \label{fig:LCC_scheme}
\end{figure}

We measure the reflection from the Fabry-Perot cavity composed of a NiO slab and a liquid crystal layer, separated by 2.5~mm (see Fig.~\ref{fig:LCC_scheme}(a)),
using a THz time-domain spectrometer (TDS), as it is shown Fig.~\ref{fig:LCC_scheme}(c). We collect the spectra as a function of the NiO temperature ($T$) and the voltage bias ($U$) applied to the liquid crystal (see the Methods section for details). Two selected time-domain traces for $U = 0$ V and $30$ V, collected at $T = 353$ K, are presented in Fig.~\ref{fig:LCC_scheme}(b), where we mark reflections from consecutive interfaces with numbers 1-5 and dashed lines (see section S1 in the Supporting Information for other raw time-domain traces). A clear voltage-induced change in the signal is observed in the second peak, corresponding to a THz pulse reflected from the liquid crystal cell.

\subsection{Temperature Dependence}

\begin{figure}[t!]
\centerline{\includegraphics[width=0.52\textwidth]{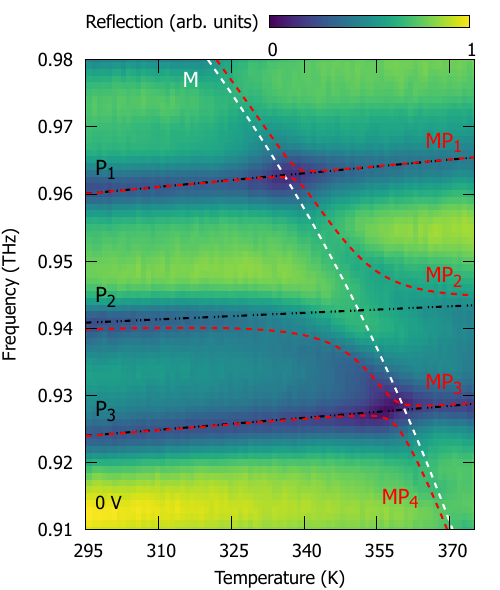}}
\caption{Reflection spectra of the Fabry-Perot cavity embedding a liquid crystal layer and an antiferromagnetic NiO layer plotted as a function of $T$ for voltage bias of $U=0$~V. The red dashed lines indicate the magnon-polariton energies obtained from the fit (see text), while the white and black dashed lines represent the energy of uncoupled magnon and consecutive Fabry-Perot cavity modes, respectively.}
\label{fig:T0V}
\end{figure}

The temperature dependence of reflection spectra of the studied Fabry-Perot cavity at the voltage bias fixed at $U = 0$~V is shown in Fig.~\ref{fig:T0V}.

To determine the coupling strength between the magnon and Fabry-Perot modes, we fit magnon-polariton energies obtained in a frame of a four-level coupled oscillator model\cite{Weisbuch:PRL1992, Sciesiek:CM2020,  Fas:JPChLett2021} to the experimental data.
In general, the coupled oscillator model assumes a coherent exchange of energy between two or more strongly coupled states, resulting in the formation of the two or more polariton branches. The Rabi splitting separating polariton branches is an analogue of a normal-mode splitting observed in the case of, e.g., coupled mechanical oscillators. Here, we assume that the magnon is coupled with three optical modes. We neglect the coupling between the optical modes. This results in the Hamiltonian describing the coupled system as shown below

\begin{equation}
  H=
  \tikz[baseline=(M.west)]{%
    \node[matrix of math nodes,matrix anchor=west,left delimiter=(,right
delimiter=),ampersand replacement=\&, minimum width=4em] (M) {%
    M           \& \Omega_1 /2 \& \Omega_2 /2                  \& \Omega_3 /2 \\
    \Omega_1 /2 \& P_1             \& 0   \& 0 \\
    \Omega_2 /2                \& 0 \& P_2 \& 0
\\
    \Omega_3 /2      \& 0      \& 0 \& P_3\\
    };
%    \node[draw,fit=(M-1-1)(M-2-2),inner sep=-1pt, rounded
corners=3pt,color=blue] {};
%    \node[draw,fit=(M-2-2)(M-3-3),inner sep=-1pt, rounded corners=3pt,
dashed] {};
%    \node[draw,fit=(M-3-3)(M-4-4),inner sep=-1pt, rounded corners=3pt,
color=red] {};
  }.
  \label{Hamiltonian}
\end{equation}
where \textit{M} denotes the energy of bare (uncoupled) magnon, \textit{P$_i$} ($i = 1$, $2$, $3$) denote energies of three selected optical modes of the Fabry-Perot cavity, and $\Omega_i$ ($i = 1$, $2$, $3$) represent the interaction strength between the magnon and the respective optical mode. The frequencies of the uncoupled magnon and Fabry-Perot cavity modes are determined from the reflectivity spectra registered at such conditions, where the detuning between the magnon and the cavity modes is much larger than their mutual coupling. The frequencies in the intermediate regions, where the coupling is meaningful, are obtained from interpolation with a polynomial function.  We fit the eigenvalues of $H$ to the minima of the $T$-dependent reflection spectra for each applied $U$ (see Fig.~\ref{fig:T0V} for the spectra at $U = 0$~V). In that way, we obtain the temperature dependencies of the polariton frequencies \textit{MP$_i$} ($i = 1$, $2$, $3$, $4$). The values obtained from the fit at 0 V are $\Omega_1$ = 2 GHz, $\Omega_2$ = 16 GHz,  $\Omega_3$ = 4 GHz.

We assume the strong coupling regime as the condition where the Rabi splitting exceeds the arithmetic average of the linewidths of uncoupled modes. We use Lorentzian to determine the linewidths of the uncoupled cavity modes, $\gamma_{Pi}$, and that of the magnon, $\gamma_M$.\cite{MBialek_AFM2024} With $(\gamma_{P2} + \gamma_M)/2 = 7.7$ GHz, the $M$–$P_2$ coupling can thus be classified as strong (see section S3 in the Supporting Information)
The interactions of the magnon with the modes \textit{P$_1$} and \textit{$P_3$} do not exhibit resolvable splittings at zero voltage bias. We attribute the strongest coupling observed for the \textit{P$_2$} mode to its largest spatial overlap with the NiO layer. Our calculations of the spatial mode distribution, presented later, confirm this interpretation.

To place our strongly coupled system in the context of previous works reporting polariton physics, we compare the obtained coupling strength to the linewidth of the polariton transition. In our case, this ratio amounts to 2.0. 
The strong coupling between the cavity mode and antiferromagnetic magnon in $\alpha$-Fe$_2$⁢O$_3$ resulted in a ratio of around 6.\cite{MBialek_PRA2021} In excitonic systems, based on GaAs, CdTe, or 2D layered semiconductors, the ratio typically remains in the range between 2 and 6.\cite{Skolnick1998, Sciesiek:CM2020, Dufferwiel2015} In systems, where plasmonic mode interacts with excitons to form a polariton, the ratio typically remains in the range of up to 5.\cite{Torma2014}

\subsection{Voltage Bias Dependence}

To demonstrate remote electric-field control of MP coupling, we apply a voltage bias to the liquid crystal cell. An electric field causes the liquid crystal molecules to reorient. At $U=0$~V, the molecules in the liquid crystal are oriented along the plane of the liquid crystal cell (the configuration known as planar alignment\cite{RKowerdziej_SR2019}), as schematically shown in Fig.~\ref{fig:LCC_scheme}(a) and in the Methods section. When the $U$ exceeds a threshold value ($U_{th}$), the molecules in the liquid crystal start to reorient continuously with the increasing voltage. At a saturation voltage ($U_s$), the molecules are aligned perpendicular to the liquid crystal cell plane, as shown in Fig.~\ref{fig:LCC_scheme}(b). The $U_{th}$ and $U_s$ depend, in general, on the liquid crystal cell construction; in our case, $U_{th} \sim 13$~V and $U_s\sim30$~V. The reorientation of the molecules causes a continuous change in the refractive index of the liquid crystal, from the initial value $n_o \sim 1.554$ at 0~V bias to $n_e \sim 1.941$ at the saturation bias. The change in the refractive index results in modification of frequencies and the electromagnetic field spatial distribution of the Fabry-Perot modes. More details on the voltage bias dependence of uncoupled cavity modes are presented in section S4 of the Supporting Information.

\begin{figure}[t!]
\centerline{\includegraphics[width=0.5\textwidth]{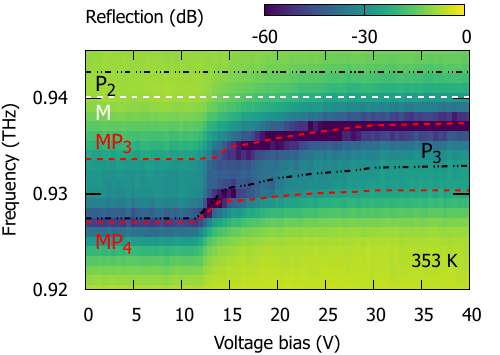}}
\caption{Reflection spectra of the Fabry-Perot cavity embedding a liquid crystal layer and an antiferromagnetic NiO layer plotted as a function of voltage bias at $T=353$~K. The red, black, and white dashed lines indicate the magnon-polariton, uncoupled FP modes, and uncoupled magnon energies obtained from the fit.}
\label{fig:U353Ka-f}
\end{figure}

We show in Fig.\ \ref{fig:U353Ka-f},
the voltage dependence of the reflection spectra of the studied Fabry–Perot cavity at a constant temperature of $T = 353$~K. Here, we focus our analysis on the interaction between the magnon and the $P_3$ mode, as its coupling strength $\Omega_3$ shows a pronounced voltage dependence, although the overall coupling to the $P_2$ mode remains considerably stronger. 
We observe, as a function of voltage, a clear anticrossing of $MP_3$ and $MP_4$ modes, which is assisted by oscillator strength transfer from $MP_4$ to $MP_3$. At around 15 V, the amplitudes of $MP_3$ and $MP_4$ are equal. This anticrossing occurs when the magnon couples simultaneously to the $P_2$ and $P_3$ modes, with the interaction involving $P_2$ being much stronger. Under these conditions, the weaker $M$–$P_3$ coupling gives rise to a small anticrossing centered at 932~GHz, while the magnon mode (940~GHz) remains strongly detuned from this resonance, as shown in Fig.\ \ref{fig:U353Ka-f}. Similar $P_2$-dominated interactions are also observed for the $P_1$ and $P_3$ modes in Fig.\ \ref{fig:T0V}. 

%%%%%%%%%%%%%%%%%%%%%%%%%%%%%

\begin{figure*}
\centerline{\includegraphics[width=1.\textwidth]{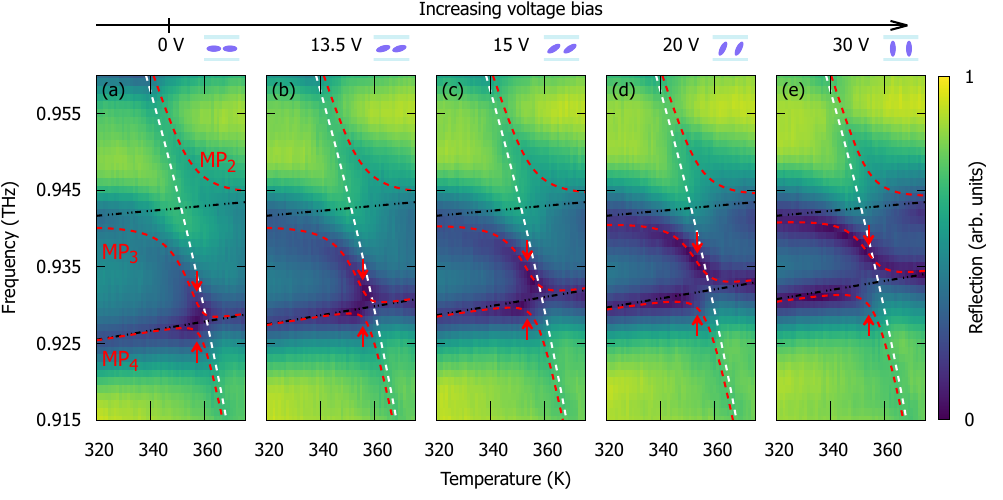}}
\caption{Temperature-dependent reflection spectra of the Fabry-Perot cavity for the voltage bias values selected in the range from $U=0$~V to $U_s=30$~V. The $U$ values are indicated above each panel, along with a graphical representation of the respective orientation of the liquid crystal molecules relative to the cell plane. The red dashed lines represent the energies of magnon-polariton branches obtained from the fitting of eigenvalues of Hamiltonian $H$ (Eq.~\ref{Hamiltonian}), while the white and black dashed lines represent the energy of the uncoupled magnon and modes of the Fabry-Perot cavity, respectively.}
\label{fig:T5U}
\end{figure*}

\begin{figure*}[t!]
\centerline{\includegraphics[width=1.\textwidth]{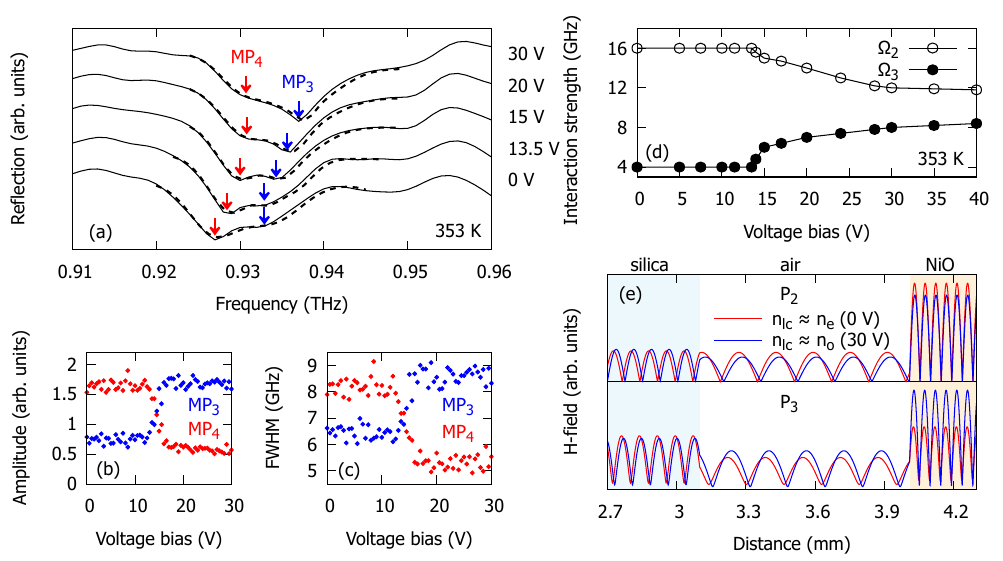}}
\caption{(a) Solid lines: reflection spectra of the Fabry-Perot cavity for selected values of voltage bias $U$ at the temperature of 353~K. The spectra are shifted vertically for clarity. The \textit{MP$_3$} and \textit{MP$_4$} transitions are indicated with blue and red arrows, respectively. Dashed lines: results of the fitting. (b) Amplitude and (c) FWHM of the \textit{MP$_3$} and \textit{MP$_4$} transitions as a function of $U$, respectively. (d) Tuning of the interaction strength between the magnon and Fabry-Perot cavity modes, $\Omega_2$ and $\Omega_3$, by the applied voltage bias. (e) $H$-field distribution of $P_2$ and $P_3$ for two selected values of $U$.}
\label{fig:U353K-L}
\end{figure*}

To better understand the voltage dependence observed in Fig.~\ref{fig:U353Ka-f}, we determine the coupling strengths $\Omega_i$ by fitting the eigenvalues of $H$ to the reflection minima as a function of $T$ for selected values of $U$ ranging from 0~V to 30~V, as shown in Fig.\ \ref{fig:T5U}.
With the increasing $U$, we observe an increase in splittings between polariton branches \textit{MP$_3$} and \textit{MP$_4$} in the vicinity of the resonance of the uncoupled $P_3$ and magnon transitions occurring at around $0.93$~THz and $353$~K. The shift of the splitting center from the resonance results from the contribution from the $P_2$ mode to the \textit{MP$_3$} and \textit{MP$_4$} wavefunctions. The increase of the splitting saturates at $U_s=30$~V. On the contrary, the increase of the $U$ from 0~V to $U_s$ leads to a decrease in the splitting between the branches \textit{MP$_2$} and \textit{MP$_3$}, indicating a reduction of the interaction strength of the magnon and the \textit{P$_2$} mode. We observe that the frequency of uncoupled $P_2$ is almost voltage-independent, while that of $P_3$ mode blueshifts by about 5~GHz with maximum bias. This difference is related to the spatial distribution of each mode in the liquid crystal layer, discussed in detail in Fig.~\ref{fig:U353K-L}(e).

In Fig.~\ref{fig:U353K-L}(a), we present the same reflection spectra as in Fig.\ \ref{fig:U353Ka-f}in the waterfall format for selected values of $U$. We see the anticrossing of \textit{MP$_3$} and \textit{MP$_4$} marked with blue and red arrows, respectively. We fitted two Gaussians to the \textit{MP$_3$} and \textit{MP$_4$} transitions to quantify their amplitudes and full widths at half maximum (FWHM). The fits show voltage-induced transfer of amplitude and width from the \textit{MP$_4$} to the \textit{MP$_3$} (see Fig.~\ref{fig:U353K-L}(b,c)).

Values of $\Omega_2$ and $\Omega_3$ as a function of $U$ at $T = 353$~K obtained from the fit are presented in Fig.~\ref{fig:U353K-L}(d). Notably, we observed a voltage-induced two-fold increase of $\Omega_3$ and a reduction of $\Omega_2$ to approximately 70\% of its value at 0~V. Importantly, these changes originate from large modifications in the spatial distribution of the electromagnetic field within the cavity, as explained in the following paragraphs. 

\subsection{Simulations}

Our transfer matrix method calculations of the mode profiles (see Fig.~\ref{fig:U353K-L}(e) and details in the Methods Simulations section) reveal that the overlap of mode P3 with the antiferromagnetic NiO layer increases by about 50\% when the voltage is increased from 0 V to 30 V, while the overlap of P2 decreases slightly from its initial value.
This behavior can be attributed to the frequencies of $P_3$ and $P_2$, which are near the frequency of one of the modes of the free-standing NiO crystal. In this case, altering the refractive index of the liquid crystal layer strongly shifts $P_3$ field amplitude within the NiO layer. This confirms that the voltage bias applied to the liquid crystal enables externally-actuated control of the spatial overlap between the cavity mode magnetic field ($H$-field) and the NiO layer, thereby modulating the magnon-photon coupling strength, without altering the underlying mode frequencies significantly.

The $P_2$ mode remains strongly coupled with the magnon independently of voltage bias. However, the $P_3$ mode, at zero bias, is weakly coupled to the magnon. It enters the strong coupling regime at the voltage bias of around 20 V, when the $\Omega_3$ exceeds the average of the linewidths of the uncoupled magnon and the $P_3$ mode.

\begin{figure}[t!]
\centerline{\includegraphics[width=0.52\textwidth]{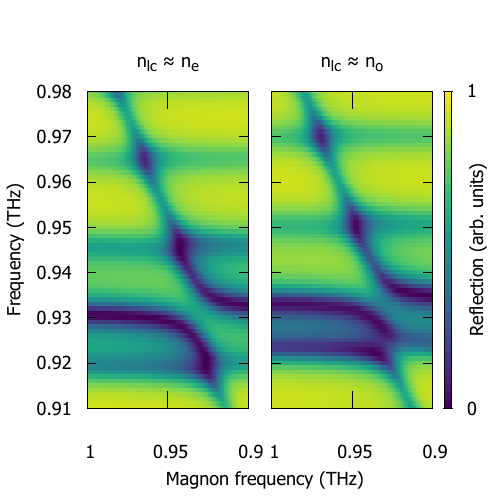}}
\caption{Reflection spectra simulated using the transfer matrix method for two values of refractive index $n_{lc}$ of the liquid crystal layer.}
\label{fig:s+m}
\end{figure}
In Fig.\ \ref{fig:s+m}, we simulate reflection spectra with the magnon, reproducing the anticrossings with two cavity modes. 
This calculation shows qualitatively similar behavior to the experiment.

\section{Conclusions}\label{sec3}

We show the remote electrical control of THz magnon-polaritons in a Fabry-Perot cavity embedding an antiferromagnetic NiO layer and a highly birefringent nematic liquid crystal. By applying the voltage to the liquid crystal layer, we tune the liquid crystal refractive index and, consequently, the spatial distribution of the Fabry-Perot cavity modes. The change of cavity mode $H$-field overlap with the NiO layer tunes the magnon-photon coupling strength and thus the vacuum Rabi splitting of magnon-polariton modes. Importantly, our results were obtained above room temperature, without external magnetic fields, and with the liquid crystal layer millimeters away from the magnetic material. Reported remote electric-field tunability is a key advancement for remote control of information transfer in the THz opto-spintronics.

\section{Methods}

\subsection{Experimental Setup}

We present the schematics of the experimental setup in Fig.~\ref{fig:LCC_scheme}(c). We used a NiO single crystal with dimensions of 5$\times$5$\times$0.33~mm$^3$, cut in the 111 plane ($MaTeck\ GmbH$). The low spin-damping rates of NiO make it an excellent material for MP studies. We tune the magnon frequency by fixing the NiO crystal on a copper plate using thermo-conductive paste, characterized by strong absorption in the THz range. The plate is placed on top of a stack of Peltier elements. The NiO temperature is varied in a range of $295$--$375$ K, with a step size of $1$ K, while the voltage range is $0$--$40$ V, with a step size of $1$ V. Crystal temperature is monitored with a $K$-type thermocouple sensor placed in the copper plate. Temperature is stabilized with a software-based PID loop controlling the current supplied to the stack of Peltier elements. The Peltier elements with the plate and NiO are placed on a kinematic mount and an x,y,z stage. The liquid crystal cell is placed on another kinematic mount, allowing precise positioning of the cell relative to the NiO crystal.  

We use a commercial TeraFlash proTHz time-domain spectrometer (TOPTICA Photonics).
This 80 MHz system employs a pair of photoconductive antennas (emitter and detector) and provides a reliable spectral bandwidth from 0.3 to 2 THz with a frequency resolution of 5 GHz (200 ps scan range) and a signal-to-noise ratio better than 120 dB.
We measure THz reflection spectra from the Fabry-Perot cavity at a 0-degree incidence angle. We use three parabolic 1" mirrors to focus the incident beam on the Fabry-Perot cavity and to collect the reflected THz pulses (Fig.\ref{fig:LCC_scheme}(c)). We use a beam splitter made of Kapton tape to direct the reflected radiation to the detector. We place the beam splitter in the parallel beam formed by the parabolic mirrors. The THz optical path is filled with dry air to minimize water-vapor absorption.

\subsection{Fourier analysis}
In the time-domain traces, we measured with a photoconductive antenna the electric field of terahertz pulses arriving at the detector. Each trace represents the temporal evolution of the electric field, including the main pulse and subsequent echoes caused by multiple reflections within the sample and between optical components. This direct measurement of the electric field (rather than intensity) preserves both amplitude and phase information, which is essential for accurate spectral analysis.
Multiple pulses observed in the time domain correspond to reflections within the sample, and their interference produces characteristic oscillations in the frequency domain. These oscillations appear as deep periodic minima in the amplitude spectrum, which are directly related to Fabry–Perot cavity modes formed between parallel interfaces of the sample.
To obtain electric field reflection spectra in the frequency domain presented in the main manuscript, we applied the fast Fourier transform to entire measured time-domain traces (200 ps, zero-padded to 1000 ps) using rectangular windowing, i.e.\ without applying any additional smoothing or apodization functions, ensuring that all reflections from multiple interfaces are included in the resulting spectra.
In the case of Fig. S5 and S6 in the Supporting Information, to filter the frequency domain spectra, we applied rectangular windowing edges set at points where the signal was near zero in time domain.
The absolute value of the obtained electric field reflection spectra is scaled to each spectrum's maximum.

\subsection{Liquid Crystal Cavity}
We present a schematic of the liquid crystal cell cross-section in Fig.~\ref{fig:LCC_scheme}(d). We use the liquid crystal cell functionalized with a highly birefringent nematic liquid crystal mixture 1825, which is characterized by ordinary and extraordinary refractive indices of $n_o = 1.554 + i \cdot 0.018$ and $n_e = 1.941 + i \cdot 0.022$, respectively\cite{RKowerdziej_APL2013, MReuter_APLM2013}. We realize a planar cell arrangement, where a 100~$\mu$m thick layer of liquid crystal is sandwiched between two pristine slabs of fused quartz (Silica), each 1500~$\mu$m thick. The surfaces of the silica slabs facing the liquid crystal are coated with a thin film of a conducting polymer, Electra 92 (AR-PC 5090.02 by $AllResist$), and uniformly rubbed with a 100~nm-thick PMMA resist. The PMMA film served as an alignment layer, facilitating the alignment of the liquid crystal molecules near its surface in the direction of rubbing. We achieve planar alignment of the liquid crystal in its bulk by matching the rubbing directions at the opposite sides of the cell.

The coating parameters for each layer are as follows: first, silica glasses are covered with Electra 92 via spin-coating (2000~rpm for 60~s) and then baked on a hotplate at 105$^\circ$C for 5 minutes. According to the product information, the resulting coated film should have a thickness of 60~nm. Next, a 100~nm layer of PMMA resist (200K, 2$\%$ by $AllResist$) is deposited on top of the Electra 92 layer by spin-coating (6000 rpm for 45~s). The resist is baked on a hotplate at 150$^\circ$C for 10~minutes. Electra is a conductive coating used in e-beam lithography, and it is semi-transparent in the THz range, so this layered structure (Electra-PMMA) forms a THz semi-transparent electrode.

Electrical contacts to the liquid crystal are made by wire soldering to the conducting polymer layer through the PMMA using tin. A signal generator with a signal amplifier supplies the liquid crystal with a sinusoidal signal at a frequency of about 80~Hz.

\subsection{Simulations}

We perform qualitative simulations of electromagnetic field distributions of the Fabry-Perot cavity modes using the transfer matrix method. We calculate the reflection from a system composed of parallel-plane slabs of silica glass, liquid crystal, NiO (without magnon resonance), and a gap filled with air between NiO and the liquid crystal cell. We neglect the thin layers of PMMA and Electra 92 in our calculations. We normalize the $H$-field for 30~V for each mode to the $H$-field at 0~V in silica glass. We assume thicknesses and refractive indices: silica 1.5~mm $n_g=2.1$, liquid crystal 0.1~mm $n_{lc}$in the range of $2.0\approx n_e$ to $1.6\approx n_o$, gap between the liquid crystal cell and NiO of 0.91 mm, and NiO thickness of 335~$\mu$m and refractive index $n_{NiO}=3.4$. In this analysis, we focus on modes that remain unaffected by interactions with the magnon mode, setting $\mu$ = 1. We use a thermocouductive paste to fix the NiO on a copper plate, which is necessary for precise temperature control. We experimentally find that the reflection spectra from NiO placed on metal, with and without paste, are very distinct. This is because the paste we use has strong absorption and relatively small reflectivity in the THz range, which we approximate in our model as a 20~$\mu$m-thick layer with $n_p=1+i5$.

In Fig.\ \ref{fig:s+m}, we simulate reflection spectra with the magnon.
Based on the NiO transmission result, we use the Lorentzian to describe the NiO magnon \cite{MBialek_AFM2024}, ie
\begin{equation}
    \mu = 1 + \frac{\Delta\mu f_m^2}{f_m^2-f^2-ifg},
\end{equation}
where $f_m$ is temperature-dependent magnon frequency, $\Delta\mu=4\cdot10^{-4}$ is its oscillator strength, and $g=8.0$ GHz is its width. We sweep $f_m$ in the range of 0.91 to 0.98 THz.
Detailed frequency shifts of modes by about 10 GHz are most likely caused by approximations in the thicknesses of certain layers and their refractive indices.

\section*{Associated Content}
%\begin{itemize}
%\item Dmitriy Yavorskiy, Jan Suffczyński, Rafał Kowerdziej, Olga Strzeżysz, Jerzy Wróbel, Wojciech Knap, Marcin Białek; Terahertz magnon-polaritons control using a tunable liquid crystal
%cavity (2025) 2504.11293 arXiv. https://arXiv:2504.11293 (accessed November 4, 2025). 

%\item 
The data that support the findings of this study are available in RepOD at https://doi.org/10.18150/WPQCOV
%\end{itemize}

%%%%%%%%%%%%%%%%%%%%%%%%%%%%%%%%%%%%%%%%%%%%%%%%%%%%%%%%%%%%%%%%%%%%%
\section*{Acknowledgements}
The authors thank Prof.\ W.\ Pacuski, and Prof.\ J.-Ph.\ Ansermet for valuable discussions.

\subsection*{Funding Sources}
The European Union supported the work through the ERC-ADVANCED grant TERAPLASM (No. 101053716). Views and opinions expressed are, however, those of the author(s) only and do not necessarily reflect those of the European Union or the European Research Council Executive Agency. Neither the European Union nor the granting authority can be held responsible for them. We acknowledge the support of the "Center for Terahertz Research and Applications (CENTERA2)" project (FENG.02.01-IP.05-T004/23) carried out within the "International Research Agendas" program of the Foundation for Polish Science, co-financed by the European Union under the European Funds for a Smart Economy Programme. This work was partially supported by Pasific2 of the Polish Academy of Sciences, sponsored by the European Union's Horizon 2020 research and innovation program under the Marie Sklodowska-Curie grant agreement No.\ 847639 and by the Ministry of Education and Science of Poland. M. B. acknowledges the financial support of the Sonata BIS-13 No. 2023/50/E/ST3/00584 grant of the National Science Centre of Poland. R. K. and O. S. acknowledge the financial support from the National Science Centre under grant SONATA BIS-12 No.\ 2022/46/E/ST7/00454.

\section*{Supporting Information}
Supporting Information is available: S1. Time-domain reflection spectra; S2. NiO magnon resonance; S3. Spectra deconvolution; S4. Voltage bias effect on cavity modes.

%\begin{suppinfo}
%The Supporting Information is available at https://pubs.acs.org/doi/...
%\end{suppinfo}
\bibliography{bib}

@article{MReuter_APLM2013,
  title = {{Highly birefringent, low-loss liquid crystals for terahertz applications}},
  author = {M. Reuter and N. Vieweg and B. M. Fischer and M. Mikulicz and M. Koch and K. Garbat and R. Dąbrowski},
  journal = {APL Materials},
  volume = {1},
  pages = {012107 },
  year = {2013},
  doi = {10.1063/1.4808244}
}

@article{AKumar_APL2011,
  title = {{Electric control of magnon frequencies and magnetic moment of bismuth ferrite thin films at room temperature}},
  author = {Ashok Kumar and J. F. Scott and and R. S. Katiyar},
  journal = {Applied Physics Letters},
  volume = {99},
  pages = {062504},
  year = {2011},
  doi = {10.1063/1.3624845}
}

@article{PRovillain_NM2010,
  title = {{Electric-field control of spin waves at room temperature in multiferroic BiFeO$_3$}},
  author = {P. Rovillain and R. de Sousa and Y. Gallais and A. Sacuto and M. A. M\'easson and D. Colson and A. Forget and M. Bibes and A. Barth\'el\'emy and M. Cazayous},
  journal = {Nature Materials},
  volume = {9},
  pages = {975–979},
  year = {2010},
  doi = {10.1038/nmat2899}
}

@article{SGuo_NL2023,
  title = {{Strong on-Chip Microwave Photon–Magnon Coupling Using Ultralow-Damping Epitaxial Y$_3$Fe$_5$O$_{12}$ Films at 2 K}},
  author = {Side Guo and Daniel Russell and Joseph Lanier and Haotian Da and P. Chris Hammel and Fengyuan Yang},
  journal = {Nano Letters},
  volume = {23},
  pages = {5055–5060},
  year = {2023},
  doi = {10.1364/OE.17.007377}
}

@article{ZShen_APLP2024,
  title = {{A liquid crystal-based multi-bit terahertz reconfigurable intelligent surface}},
  author = {Ze Shen and Weili Li and Biaobing Jin and Dixian Zhao},
  journal = {APL Photonics},
  volume = {9},
  pages = {016109},
  year = {2024},
  doi = {10.1063/5.0176272}
}

@article{LWang_LSA2015,
  title = {{Broadband tunable liquid crystal terahertz waveplates
driven with porous graphene electrodes}},
  author = {Lei Wang and Xiao-Wen Lin and Wei Hu and Guang-Hao Shao and Peng Chen and Lan-Ju Liang and Biao-Bing Jin and Pei-Heng Wu and Hao Qian and Yi-Nong Lu and Xiao Liang and Zhi-Gang Zheng and Yan-Qing Lu},
  journal = {Light: Science $\&$ Applications},
  volume = {4},
  pages = {1-6},
  year = {2015},
  doi = {10.1038/lsa.2015.26}
}

@article{DShrekenhamer_PRL2013,
  title = {{Liquid Crystal Tunable Metamaterial Absorber}},
  author = {David Shrekenhamer and Wen-Chen Chen and Willie J. Padilla},
  journal = {Physical Review Letters},
  volume = {110},
  pages = {177403},
  year = {2013},
  doi = {10.1103/PhysRevLett.110.177403}
}

@article{MBialek_AFM2024,
  title = {{Hybridization of Terahertz Phonons and Magnons in Disparate and Spatially-Separated Material Specimens}},
  author = {M. Białek and Y. Todorov and K. Stelmaszczyk and D. Szwagierczak and B. Synkiewicz-Musialska and J. Kulawik and N. Pałka and M. Potemski and W. Knap},
  journal = {Advanced Functional Materials},
  volume = {},
  pages = {2416037},
  year = {2024},
  doi = {10.1002/adfm.202416037},
  url = {https://doi.org/10.1002/adfm.202416037}
}

@article{MBialek_APL2022,
  title = {{Antiferromagnetic resonance in $\alpha$-Fe$_2$O$_3$ up to its Néel temperature}},
  author = {M. Białek and J. Zhang and H. Yu and J.-Ph. Ansermet},
  journal = {Applied Physics Letters},
  volume = {121},
  pages = {032401},
  year = {2022},
  doi = {10.1063/5.0094868},
  url = {https://doi.org/10.1063/5.0094868}
}

@article{PSivarajah_JAP2019,
  title = {{THz-frequency magnon-phonon-polaritons in the collective strong-coupling regime}},
  author = {Prasahnt Sivarajah and Andreas Steinbacher and Blake Dastrup and Jian Lu and Maolin Xiang and Wei Ren and Stanislav Kamba and Shixun Cao and Keith A. Nelson},
  journal = {Journal of Applied Physics},
  volume = {125},
  pages = {213103},
  year = {2019},
  doi = {10.1063/1.5083849},
  url = {https://doi.org/10.1063/1.5083849}
}

@article{RKowerdziej_LC2016,
  title = {Active control of terahertz radiation using a metamaterial loaded with a nematic liquid crystal},
  author = {Rafał Kowerdziej and Leszek Jaroszewicz},
  journal = {Liquid Crystals},
  volume = {43},
  pages = {1120-1125},
  year = {2016},
  doi = {10.1080/02678292.2016.1160297},
  url = {https://doi.org/10.1080/02678292.2016.1160297}
}

@article{OBuchnev_APL2013,
  title = {Controlling intensity and phase of terahertz radiation with an optically thin liquid crystal-loaded metamaterial},
  author = {O. Buchnev and J. Wallauer and M. Walther and M. Kaczmarek and N. I. Zheludev and V. A. Fedotov},
  journal = {Applied Physics Letters},
  volume = {103},
  pages = {141904},
  numpages = {141904},
  year = {2013},
  doi = {10.1063/1.4823822},
  url = {https://doi.org/10.1063/1.4823822}
}

@article{NChikhi_SR2016,
  title = {A hybrid tunable THz metadevice using a high birefringence liquid
crystal},
  author = {Nassim Chikhi and Mikhail Lisitskiy and Gianpaolo Papari and Volodymyr Tkachenko and Antonello Andreone},
  journal = {Scientific Reports},
  volume = {6},
  pages = {34536},
  numpages = {34536},
  year = {2016},
  doi = {10.1038/srep34536},
  url = {https://doi.org/10.1038/srep34536}
}

@article{GDENG_OME2021,
  title = {Tunable terahertz metamaterial wideband absorber with liquid crystal},
  author = {GUANGSHENG Deng and HUALONG Hu and HAISHENG Mo and JUNJIE
Xu and ZHIPING Yin and HONGBO Lu and MINGGANG Hu and JIAN Li and 
JUN Yang},
  journal = {Optical Materials Express},
  volume = {11},
  pages = {4026-4035},
  numpages = {4026-4035},
  year = {2021},
  doi = {10.1364/OME.444899},
  url = {https://doi.org/10.1364/OME.444899}
}

@article{Abe_APL2011,
author = {Abe, E.  and Wu, H  and Ardavan, A.  and Morton,J. J. L. },
title = {Electron spin ensemble strongly coupled to a three-dimensional microwave cavity},
journal = {Applied Physics Letters},
volume = {98},
pages = {251108},
year = {2011},
doi = {10.1063/1.3601930},
URL = {https://doi.org/10.1063/1.3601930
}
}

@article{Zhang_PRL2014,
  title = {Strongly Coupled Magnons and Cavity Microwave Photons},
  author = {Zhang, X. and Zou, Ch.-L. and Jiang, L. and Tang, H. X.},
  journal = {Phys. Rev. Lett.},
  volume = {113},
  pages = {156401},
  numpages = {5},
  year = {2014},
  month = {Oct},
  publisher = {American Physical Society},
  doi = {10.1103/PhysRevLett.113.156401},
  url = {https://link.aps.org/doi/10.1103/PhysRevLett.113.156401}
}

@article{Tabuchi_PRL2014,
  title = {Hybridizing Ferromagnetic Magnons and Microwave Photons in the Quantum Limit},
  author = {Tabuchi, Y. and Ishino, S. and Ishikawa, T. and Yamazaki, R. and Usami, K. and Nakamura, Y.},
  journal = {Phys. Rev. Lett.},
  volume = {113},
  pages = {083603},
  numpages = {5},
  year = {2014},
  month = {Aug},
  publisher = {American Physical Society},
  doi = {10.1103/PhysRevLett.113.083603}
}

@Article{Zhang_NC2015,
author={Zhang, X.
and Zou, Ch.-L.
and Zhu, Na
and Marquardt, F.
and Jiang, L.
and Tang, H. X.},
title={Magnon dark modes and gradient memory},
journal={Nature Communications},
year={2015},
month={Nov},
day={16},
volume={6},
pages={8914},
issn={2041-1723},
doi={10.1038/ncomms9914},
url={https://doi.org/10.1038/ncomms9914}
}

@article{Everts_PRB2020,
  title = {Ultrastrong coupling between a microwave resonator and antiferromagnetic resonances of rare-earth ion spins},
  author = {Everts, J. R. and King, G. G. G. and Lambert, N. J. and Kocsis, S. and Rogge, S. and Longdell, J. J.},
  journal = {Phys. Rev. B},
  volume = {101},
  pages = {214414},
  numpages = {6},
  year = {2020},
  month = {Jun},
  publisher = {American Physical Society},
  doi = {10.1103/PhysRevB.101.214414},
  url = {https://link.aps.org/doi/10.1103/PhysRevB.101.214414}
}

@article{Potts_APL2020,
author = {Potts,C. A.  and Davis,J. P. },
title = {Strong magnon–photon coupling within a tunable cryogenic microwave cavity},
journal = {Applied Physics Letters},
volume = {116},
pages = {263503},
year = {2020},
doi = {10.1063/5.0015660},
URL = {https://doi.org/10.1063/5.0015660}
}

@article {Lachance-Quirion_S2020,
	author = {Lachance-Quirion, D. and Wolski, S. P. and Tabuchi, Y. and Kono, Sh. and Usami, K. and Nakamura, Y.},
	title = {Entanglement-based single-shot detection of a single magnon with a superconducting qubit},
	volume = {367},
	pages = {425--428},
	year = {2020},
	doi = {10.1126/science.aaz9236},
	publisher = {American Association for the Advancement of Science},
	issn = {0036-8075},
	URL = {https://science.sciencemag.org/content/367/6476/425},
	journal = {Science}
}

@article{Li_JAP2020,
author = {Li,Y.  and Zhang,W.  and Tyberkevych,V.  and Kwok,W.-K.  and Hoffmann,A.  and Novosad,V. },
title = {Hybrid magnonics: Physics, circuits, and applications for coherent information processing},
journal = {Journal of Applied Physics},
volume = {128},
pages = {130902},
year = {2020},
doi = {10.1063/5.0020277},
URL = {https://doi.org/10.1063/5.0020277}
}

@article{Bhoi_JAP2021,
author = {Bhoi,B. and Jang,S.-H.  and Kim,B.  and Kim,S.-K. },
title = {Broadband photon–magnon coupling using arrays of photon resonators},
journal = {Journal of Applied Physics},
volume = {129},
pages = {083904},
year = {2021},
doi = {10.1063/5.0040194},
URL = {https://doi.org/10.1063/5.0040194}
}

@article{KRoux_NC2020,
  title = {Strongly correlated Fermions strongly coupled to light},
  author = {Kevin Roux and Hideki Konishi and Victor Helson and Jean-Philippe Brantut},
  journal = {Nature Communications},
  volume = {11},
  pages = {2974 },
  year = {2020},
  doi = {10.1038/s41467-020-16767-8},
  url = {https://doi.org/10.1038/s41467-020-16767-8}
}

@article{HYuan_PR2022,
  title = {Quantum magnonics: When magnon spintronics meets
quantum information science},
  author = {H.Y. Yuan and Yunshan Cao and Akashdeep Kamra and Rembert A. Duine and Peng Yan},
  journal = {Physics Reports},
  volume = {965},
  pages = {1-74},
  numpages = {},
  year = {2022},
  month = {},
  publisher = {Elsevier},
  doi = {10.1016/j.physrep.2022.03.002},
  url = {https://doi.org/10.1016/j.physrep.2022.03.002}
}

@article{TElijahKritzell_AOM2023,
  title = {Terahertz Cavity Magnon Polaritons},
  author = {T. Elijah Kritzell and Andrey Baydin and Fuyang Tay and Rodolfo Rodriguez and Jacques Doumani and Hiroyuki Nojiri and Henry O. Everitt and Igor Barsukov and Junichiro Kono},
  journal = {Advanced Optical Materials},
  volume = {12},
  pages = {2302270},
  year = {2024},
  publisher = {Wiley-VCH GmbH},
  doi = {10.1002/adom.202302270},
  url = {https://doi.org/10.1002/adom.202302270}
}

@article{MBialek_PRA2021,
author = {M. Białek and J. Zhang H. Yu and J.-Ph. Ansermet},
title = {Strong Coupling of Antiferromagnetic Resonance with Subterahertz Cavity Fields},
journal = {Physical Review Applied},
volume = {15},
pages = {044018},
year = {2021},
doi = {10.1103/PhysRevApplied.15.044018},
}

@article{MBialek_PRA2023,
author = {M. Białek and W. Knap and J.-P. Ansermet},
title = {Cavity-Mediated Coupling of Terahertz Antiferromagnetic Resonators},
journal = {Physical Review Applied},
volume = {19},
pages = {064007},
year = {2023},
doi = {10.1103/PhysRevApplied.19.064007},
}

@article{MBialek_PRB2020,
author = {M. Białek and A. Magrez and J.-Ph. Ansermet},
title = {Spin-wave coupling to electromagnetic cavity fields in dysposium ferrite},
journal = {Physical Review B},
volume = {101},
pages = {024405},
year = {2020},
doi = {10.1103/PhysRevB.101.024405},
}

@article{RKowerdziej_APL2013,
author = {Rafał Kowerdziej and Marek Olifierczuk and Janusz Parka and Jerzy Wróbel},
title = {Dielectric properties of highly anisotropic nematic liquid crystals for tunable microwave components},
journal = {Applied Physics Letters},
volume = {103},
pages = {172902},
year = {2013},
doi = {10.1063/1.4826504},
}

@article{RKowerdziej_APL2014,
author = {Rafał Kowerdziej and Marek Olifierczuk and Janusz Parka and Jerzy Wróbel},
title = {Terahertz characterization of tunable metamaterial based on electrically controlled nematic liquid crystal},
journal = {Applied Physics Letters},
volume = {105},
pages = {022908},
year = {2014},
doi = {10.1063/1.4890850},
}

@article{RKowerdziej_APL2015,
author = {Rafał Kowerdziej and Leszek Jaroszewicz and Marek Olifierczuk and Janusz Parka},
title = {Experimental study on terahertz metamaterial embedded in nematic liquid crystal},
journal = {Applied Physics Letters},
volume = {106},
pages = {092905},
year = {2015},
doi = {10.1063/1.4914032},
}

@article{RKowerdziej_SR2019,
  title = {Ultrafast electrical switching of nanostructured metadevice with dual-frequency liquid crystal},
  author = {Rafał Kowerdziej and Jerzy Wróbel and Przemysław Kula},
  journal = {Scientific Reports},
  volume = {9},
  pages = {20367},
  year = {2019},
  doi = {10.1038/s41598-019-55656-z},
  url = {https://doi.org/10.1038/s41598-019-55656-z}
}

@article{XFu_ACS2022,
  title = {Flexible Terahertz Beam Manipulations Based on Liquid-Crystal-Integrated Programmable Metasurfaces},
  author = {Xiaojian Fu and Lei Shi and Jun Yang and Yuan Fu and Chenxi Liu and Jun Wei Wu and Fei Yang and Lei Bao and and Tie Jun Cui},
  journal = {ACS Applied Materials \& Interfaces},
  volume = {14},
  pages = {22287–22294},
  year = {2022},
  doi = {10.1021/acsami.2c02601},
  url = {https://doi.org/10.1021/acsami.2c02601}
}

@article{JTHou_APL2024,
  title = {Electrical manipulation of dissipation in microwave photon–magnon hybrid system through the spin Hall effect},
  author = {Justin T. Hou and Chung-Tao Chou and Jiahao Han and Yabin Fan and Luqiao Liu},
  journal = {Applied Physics Letters},
  volume = {125},
  pages = {072401},
  year = {2024},
  doi = {10.1063/5.0182270},
  url = {https://doi.org/10.1063/5.0182270}
}

@article{TGHBlank_APL2023,
  title = {Magneto-optical detection of terahertz cavity magnon-polaritons in antiferromagnetic HoFeO$_3$},
  author = {T. G. H. Blank and K. A. Grishunin and A. V. Kimel},
  journal = {Applied Physics Letters},
  volume = {122},
  pages = {072402},
  year = {2023},
  doi = {10.1063/5.0124503},
  url = {https://doi.org/10.1063/5.0124503}
}

@article{KGrishunin_ACSPh2018,
  title = {Terahertz Magnon-Polaritons in TmFeO$_3$},
  author = {Kirill Grishunin and Thomas Huisman and Guanqiao Li and Elena Mishina and Theo Rasing and Alexey V. Kimel and Kailing Zhang and Zuanming Jin and Shixun Cao and Wei Ren and Guo-Hong Ma and Rostislav V. Mikhaylovskiy},
  journal = {ACS Photonics},
  volume = {5},
  pages = {1375−1380},
  year = {2018},
  doi = {10.1021/acsphotonics.7b01402},
  url = {https://doi.org/10.1021/acsphotonics.7b01402}
}

@article{Fas:JPChLett2021,
	author = {F{\k{a}}s, Tomasz and {\ifmmode\acute{S}\else\'{S}\fi}ciesiek, Maciej and Pacuski, Wojciech and Golnik, Andrzej and Suffczy{\ifmmode\acute{n}\else\'{n}\fi}ski, Jan},
	title = {{Hybrid Semimagnetic Polaritons in a Strongly Coupled Optical Microcavity}},
	journal = {The Journal of Physical Chemistry Letters},
	volume = {12},
	pages = {7619--7624},
	year = {2021},
	month = aug,
	publisher = {American Chemical Society},
	doi = {10.1021/acs.jpclett.1c01894}
}

@article{Sciesiek:CM2020,
	author = {{\ifmmode\acute{S}\else\'{S}\fi}ciesiek, Maciej and Sawicki, Krzysztof and Pacuski, Wojciech and Sobczak, Kamil and Kazimierczuk, Tomasz and Golnik, Andrzej and Suffczy{\ifmmode\acute{n}\else\'{n}\fi}ski, Jan},
	title = {{Long-distance coupling and energy transfer between exciton states in magnetically controlled microcavities}},
	journal = {Commun. Mater.},
	volume = {1},
	number = {78},
	pages = {1--8},
	year = {2020},
	month = oct,
	issn = {2662-4443},
	publisher = {Nature Publishing Group},
	doi = {10.1038/s43246-020-00079-x}
}

@article{Kaur:APL2016,
	author = {Kaur, S. and Yao, B. M. and Rao, J. W. and Gui, Y. S. and Hu, C.-M.},
	title = {{Voltage control of cavity magnon polariton}},
	journal = {Appl. Phys. Lett.},
	volume = {109},
	number = {3},
	pages = {032404},
	year = {2016},
	month = jul,
	issn = {0003-6951},
	publisher = {AIP Publishing},
	doi = {10.1063/1.4959140}
}

@article{Rao:APL2018,
    author = {Rao, J. W. and Yao, B. M. and Fan, X. L. and Xue, D. S. and Gui, Y. S. and Hu, C.-M.},
    title = {Electric control of cooperative polariton dynamics in a cavity-magnon system},
    journal = {Applied Physics Letters},
    volume = {112},
    number = {26},
    pages = {262401},
    year = {2018},
    month = {06},
    issn = {0003-6951},
    doi = {10.1063/1.5024336},
    url = {https://doi.org/10.1063/1.5024336},
    eprint = {https://pubs.aip.org/aip/apl/article-pdf/doi/10.1063/1.5024336/19779570/262401\_1\_online.pdf},
}

@article{JChen:APL2025,
	author = {Junyu Chen and Qixin Li and Zhichao Fu and Jiamin Shang and Peng Suo  and Xian Lin and Jianlin Luo and Xinbo Wang Anhua Wu and Guohong Ma},
	title = {{Terahertz cavity magnon-polaritons in Gd$_{0.5}$Ho$_{0.5}$FeO$_3$ single crystals tuned with temperature and magnetic field}},
	journal = {Appl. Phys. Lett.},
	volume = {127},
	issue = {6},
	pages = {062402},
	year = {2025},
	month = august,
	issn = {0003-6951},
	publisher = {AIP Publishing},
	doi = {https://doi.org/10.1063/5.0277632}
}

@article{ABaydin:PRR2023,
  title = {Magnetically tuned continuous transition from weak to strong coupling in terahertz magnon polaritons},
  author = {Baydin, Andrey and Hayashida, Kenji and Makihara, Takuma and Tay, Fuyang and Ma, Xiaoxuan and Ren, Wei and Ma, Guohong and Noe, G. Timothy and Katayama, Ikufumi and Takeda, Jun and Nojiri, Hiroyuki and Cao, Shixun and Bamba, Motoaki and Kono, Junichiro},
  journal = {Phys. Rev. Res.},
  volume = {5},
  issue = {1},
  pages = {L012039},
  numpages = {6},
  year = {2023},
  month = {Mar},
  publisher = {American Physical Society},
  doi = {10.1103/PhysRevResearch.5.L012039},
  url = {https://link.aps.org/doi/10.1103/PhysRevResearch.5.L012039}
}

@article{Weisbuch:PRL1992,
  title = {Observation of the coupled exciton-photon mode splitting in a semiconductor quantum microcavity},
  author = {Weisbuch, C. and Nishioka, M. and Ishikawa, A. and Arakawa, Y.},
  journal = {Phys. Rev. Lett.},
  volume = {69},
  issue = {23},
  pages = {3314--3317},
  numpages = {0},
  year = {1992},
  month = {Dec},
  publisher = {American Physical Society},
  doi = {10.1103/PhysRevLett.69.3314},
  url = {https://link.aps.org/doi/10.1103/PhysRevLett.69.3314}
}

@article{Dufferwiel2015,
    author = {Dufferwiel, S. and Schwarz, S. and Withers, F. and Trichet, A. A. P. and Li, F. and Sich, M. and Del Pozo-Zamudio, O. and Clark, C. and Nalitov, A. and Solnyshkov, D. D. and Malpuech, G. and Novoselov, K. S. and Smith, J. M. and Skolnick, M. S. and Krizhanovskii, D. N. and Tartakovskii, A. I.},
    title = {{Exciton{\textendash}polaritons in van der Waals heterostructures embedded in tunable microcavities}},
    journal = {Nat. Commun.},
    volume = {6},
    number = {8579},
    pages = {1--7},
    year = {2015},
    month = oct,
    issn = {2041-1723},
    publisher = {Nature Publishing Group},
    doi = {10.1038/ncomms9579}
}

@article{Skolnick1998,
    author = {Skolnick, M. S. and Fisher, T. A. and Whittaker, D. M.},
    title = {{Strong coupling phenomena in quantum microcavity structures}},
    journal = {Semicond. Sci. Technol.},
    volume = {13},
    number = {7},
    pages = {645},
    year = {1998},
    month = jul,
    issn = {0268-1242},
    publisher = {IOP Publishing},
    doi = {10.1088/0268-1242/13/7/003}
}

@article{Torma2014,
    author = {T{\ifmmode\ddot{o}\else\"{o}\fi}rm{\ifmmode\ddot{a}\else\"{a}\fi}, P. and Barnes, W. L.},
    title = {{Strong coupling between surface plasmon polaritons and emitters: a review}},
    journal = {Rep. Prog. Phys.},
    volume = {78},
    number = {1},
    pages = {013901},
    year = {2014},
    month = dec,
    issn = {0034-4885},
    publisher = {IOP Publishing},
    doi = {10.1088/0034-4885/78/1/013901}
}
%%%% SUPPLEMENTARY INFORMATION %%%%

%\newpage
%\begin{suppinfo}

%\end{suppinfo}
\end{document}